\documentclass[letterpaper]{article}
\usepackage{aaai}
\usepackage{times}
\usepackage{helvet}
\usepackage{courier}
\usepackage{caption}
\usepackage{subcaption}
\usepackage{graphicx}

\begin{document}
%
\title{Dynamic Algorithmic Service Agreements Perspective}
\author{
Bogdana Rakova \\
Responsible AI, Accenture \\ 
415 Mission St Floor 35\\
San Francisco, CA 94105\\
bogdana.rakova@accenture.com \\
\And
Laura Kahn\\
Accenture Federal Services\\
800 N. Glebe Road \\
Arlington, VA 22203\\
laura.kahn@accenturefederal.com
}
\maketitle

\begin{abstract}
A multi-disciplinary understanding of the concepts of identity, agency, relationships, interactions, and information could help us develop mitigation strategies for responsible human-algorithmic systems in the field of AI. It is imperative for us to question the use of the Terms of Service (ToS) agreements model in the context of algorithmic systems, specifically AI systems that make decisions which affect people and their livelihoods. In this position paper, we identify five areas of concern in traditional ToS agreements by drawing on studies of sociotechnical systems in Science and Technology Studies: accommodating and enabling change, co-constitution, reflective directionality, friction, and generativity. We aim to address these ToS shortcomings and propose components of a novel Dynamic Algorithmic Service Agreements (DASA) framework. The DASA could be employed as a self-regulation framework while also enabling additional feedback loops between people and algorithmic systems. Rich interaction frameworks could enable us to better negotiate and cooperate with AI systems towards accomplishing the real-world goals we use them for.  We illustrate the DASA framework in the context of a Recommender System used in the curation of real and synthetic data. We do not intend for the DASA framework to replace the ToS model, but instead think it will provide practitioners with an alternative point of view for the design of dynamic interaction interfaces for AI systems that account for human identity and agency.
\end{abstract}

\section{Introduction}
Self-determination theory considers the innate psychological needs of humans for competence, autonomy, and relatedness \cite{deci2004handbook}. When designing algorithmic sociotechnical systems, maintaining a reasonable degree of human autonomy is critical to protecting the rights and liberties of the human actors who enter into interactions with them. Human autonomy is a key component of any \textit{human-in-the-loop} system. Algorithms increasingly play a critical role in curating information and automating decision-making that affects humans in a variety of sociotechnical systems, ranging from healthcare, finance, housing, employment, entertainment, and others. However, increasingly algorithms operate in relatively closed-loop systems that lack transparency to the users they affect \cite{ajunwa2019platforms,inbook,doi:10.1111/soc4.12493}. Furthermore, there are few recourse for humans to offer feedback in the systems that make recommendations and often time irreversible decisions for and about them. 

Responsible AI considers human values and ethics along the whole pipeline of design, development, evaluation, and deployment of algorithmic systems \cite{cowls2018prolegomena,8395130,askell2019role}. One key outcome of the use of Responsible AI frameworks aims to be the improved alignment of the impacts of technology and innovation with societal demands and values. Having legally binding DASAs that can accommodate for human autonomy while reflecting societal values, could enable us to empower the human actors entangled within the algorithmic sociotechnical systems we are part of \cite{latour2014better,bruni2007reassembling}. 

The main contribution of our work is to (1) outline five types of sociotechnical concerns of traditional ToS agreements in the context of AI Systems, (2) propose a Dynamic Algorithmic Service (DAS) Agreements framework, and (3) demonstrate how it could be applied in the context of a Recommender System AI models used in entertainment, advertising, and other media contexts.

\section{Related Work}
User rights such as the right to be forgotten \cite{rightToBeForgotten}, right to identity \cite{rightToIdentity}, right to reasonable inference \cite{wachter2019right}, right to explanation \cite{10.1093/idpl/ipx005}, and others, have been explored in the context of algorithmic systems. We build on that work in the context of end-user and source code license agreements that operate on the level of human-algorithmic interactions. One such agreement is the Terms of Service agreement. 

Algorithmic Contracts as well as Algorithmic Social Contracts \cite{Rahwan2017SocietyintheloopPT} have previously been proposed by legal, computer science, and other scholars. Furthermore, we see a need for the development of contracts which operate on the level of interactions between individuals and algorithmic systems. Such finer-grained contracts could then be adopted as part of a \textit{society-in-the-loop social contract} \cite{Rahwan2017SocietyintheloopPT}, however they may also exist as stand-alone agreements which are adopted within a specific community, collective, population, etc. The work on \textit{human-in-the-loop} system design within Human Factors and Ergonomics has had significant impact on the developments in the field of Human-Computer Interaction. The concept of \textit{human-in-the-loop} systems has emerged through the field of Human Supervisory Control as a process by which "one or more human operators are intermittently programming and continually receiving information from a computer that itself closes an autonomous control loop through artificial effectors to the controlled process or task environment" \cite{sheridan1992telerobotics}. As explored by Sheridan, some of the potential societal negatives of AI technology could be summarized with a single word: alienation \cite{doi:10.1002/0470048204.ch38}. We go on to pose the question of, what would it look like to employ a design framework incorporating a \textit{community-in-the-loop} approach that goes beyond the limitations of the power imbalance between a human and an algorithmic system \cite{elish2019moral,hoffmann2019fairness}. In what follows, we demonstrate the need and the opportunity to go beyond the strive towards generalization of AI systems. Instead, different communities could have the opportunity to integrate AI Systems in their collective livelihood in ways which are aligned, sustainable, and not necessarily generalizable to Society as a whole. 

To look for regulatory strategies, we take inspiration from Social Science and the field of Action Research - the iterative process of collaboration with community partners in order to deeply understand the local context of existing problems, perspectives, and challenges. The goal is not necessarily to create a generalized solution but instead to create sustainable change through the design of sociotechnical interventions that are conducted collaboratively with all involved stakeholders \cite{hayes2011relationship,gaffney2008participatory}. We build on the work of scholars in the Action Research field to propose a dynamic evaluation approach, where different levels of collectives of people have the agency to participate in the way AI Systems are being evaluated.

Drawing from the fields of Critical Algorithm Studies, Science and Technology Studies, as well as the Fairness, Accountability, and Transparency of ML community, we aim to evolve prior analysis of data and algorithmic impact assessments \cite{gebru2018datasheets,stark2019data,doi:10.1177/2053951717738104,selbst2019fairness,mitchell2019model,doi:10.1177/0163443716643157,ainow_aia,hoffmann2019fairness} by introducing an algorithmic service agreement framework.  Similarly, we build on prior work developed by the Canadian Government - an Algorithmic Impact Assessment questionnaire \cite{canada_aia} designed to assess and mitigate the risks associated with deploying an automated decision system. The assessment provides transparency into the process since it is freely available for everyone to see and reuse in other contexts. Learning from this work, we aim to demonstrate the value of dynamic algorithmic assessments in addition to qualitative questionnaires.

Our interactions with algorithmic systems lead us to develop relationships which extend beyond the digital realms. Inevitably, they would also have first and second order implications on our innate sense of agency, autonomy, and identity. We build on prior work on self-sovereignty and generative identity \cite{sheldrake_2019,verifysocial,zittrain2005generative} in order to develop a framework which encompasses algorithmic inferences.

\section{A critical look at ToS Agreements}
We start by outlining five kinds of sociotechnical concerns of ToS agreements related to AI systems used in the context of collecting and processing data as well as the production of algorithmic outcomes which are then interpreted within a social context.

\subsection{Accommodating and enabling change}
We argue that ToS agreements need to allow for mutability. From a Social Science perspective, mutability is closely connected to our ability to change, our dynamic and multifaceted human preferences, as well as multiplicity of identities \cite{rakova2019human}. AI system creators need to be cognizant of the mutability of the data variables which are used in the algorithmic decision-making process and allow for that change to happen. Furthermore, ToS agreements as well as the evaluation frameworks which are employed by AI System creators, need to be able to adapt to the dynamic nature of human identities. 


\subsection{Co-constitution}
Currently, ToS agreements do not encompass human-algorithmic interactions nor are they designed for end-user feedback. The challenge is to adequately define an agreement ex-ante without knowing what the outcome is going to be. For example, a person may not know they are discriminated against by an algorithmic system unless it has already happened. 

\subsection{Reflexive directionality}
ToS agreements fail to capture the social relationship between data and algorithmic outcomes. How does certain data lead to specific outcomes? Given the algorithmic outcome is it possible to reconstruct the input data? The fields of explainability and interpretability in AI are going to help enable such reflexive directionality, however from the perspective of a licensing framework, we may not need to understand a black-box system in order to set up a methodology which protects the rights and liberties of the users of the black-box system. Similarly to the way we don't need a neuroscience degree in order to develop a trust relationship, through understanding the mental models and decision making processes of a worker in a grocery store, when we had asked them for their help with finding a specific item. 

\subsection{Allow for friction}

Similarly to mutability, currently ToS agreements do not allow for friction. Users of a certain platform are not always notified when a change in the ToS has happened. Even if users receive notifications of changes in the ToS, often times they don't have any opportunity to express their preferences to the ToS on a more granular level. Furthermore, they do not have a way to internalize what could be the impact of such change and often do not have the autonomy to act upon it, outside of choosing to completely opt out of the platform. What if the platform allowed us to separate the claims we make about ourselves from the claims other human and algorithmic actors have made about us?

\subsection{Generativity}

Currently ToS agreements may differ from one country to another but they often fail to address the unique needs of cities, collectives, and individual communities. We define communities broadly as "an umbrella term [defined]... in geographic terms... as a neighborhood or town (place-based or communities of place definitions); or in social terms, such as a group of people sharing common chat rooms on the internet, a national professional association, or a labor union (communities of interest definitions)" \cite{community_2015}. 

Through addressing these five sociotechical concerns, we hope that our work will positively contribute towards Responsible AI systems which are better aligned with the values of individuals and communities. 

\section{A Dynamic Framework}
The framework proposed here aims to address the concerns outlined in the previous section. Acknowledging the complexity of this topic, we highlight that this perspective is work in progress which we hope will spark meaningful and tangible discussions. We pose that, the DASA model could be used as a regulatory framework while also serving as a way for people to provide additional feedback to an algorithmic system. DASA-enabled feedback loops may enable users to negotiate and cooperate with an AI system through intentionally \textit{nudging} it into a better alignment with their respective multi-faceted human identity \cite{Matias,MatiasPersuading}. 


\begin{figure}[htbp]
    \centering
    \includegraphics[width=\linewidth]{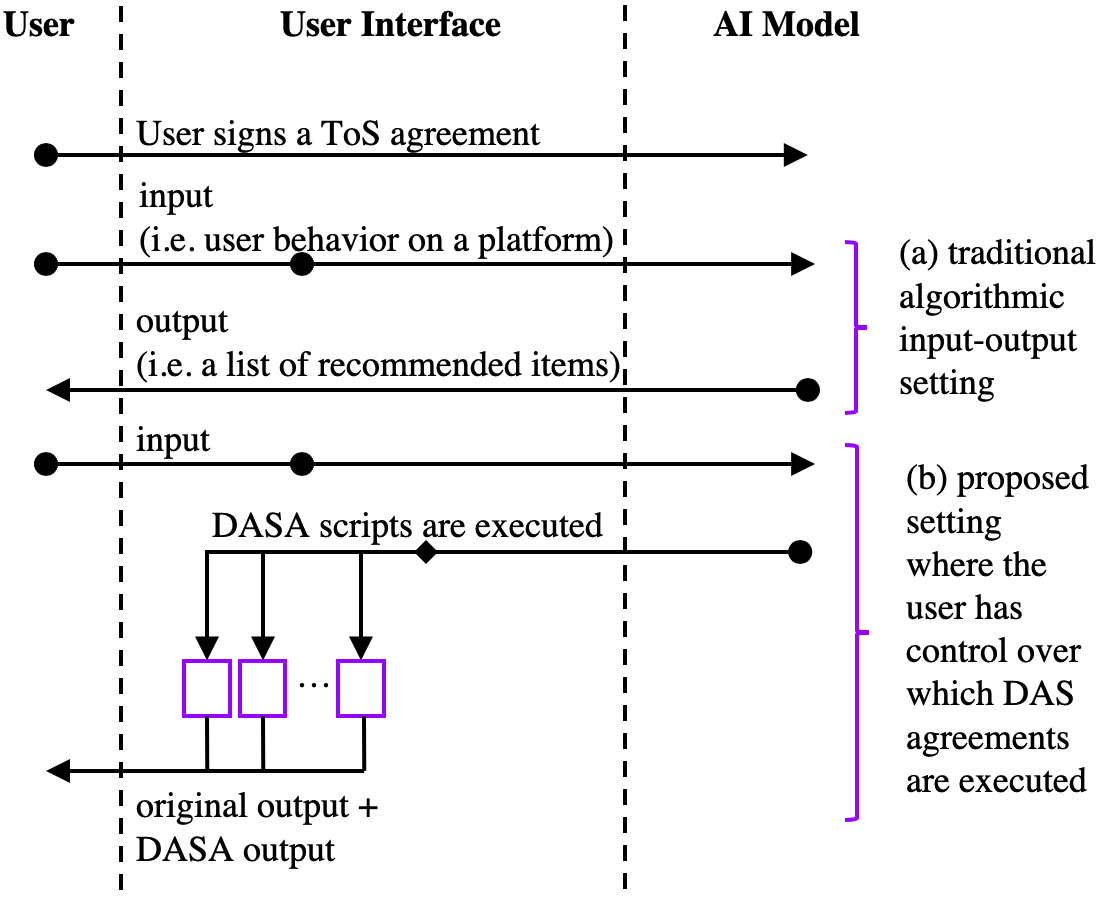}
    \caption{Outline of a traditional interaction between a user and an AI system (a) vs (b) the proposed dynamic algorithmic service agreement process diagram. In (b) users are able to select from a list of dynamic algorithmic service agreements and modify their parameters, etc. The DASA scripts are executed on each algorithmic outcome provided by the AI system. Ultimately, we highlight the need for user interfaces which allow for verification and negotiation.}
    \label{fig:setup_figure}
\end{figure}

The DASA is a set of code scripts which are hosted on an open source platform. Each script takes as its input the output of an algorithmic system and is designed to verify certain properties of the algorithmic outcome. The concrete properties we aim to verify will depend on the concrete use-case, however the general scope of the verification aims to make sure that we have algorithmic procedures which can address the sociotechnical concerns described in the previous section. The DASA scripts for a particular AI system could be created by the organization which is designing, developing, and deploying the AI system itself as well as by third party groups or organizations. The work on creating verification scripts could become a requirement or a social norm in the development of Machine Learning algorithms, similarly to the way testing has become an inseparable part of software development.

Traditionally, in Machine Learning research, there's a difference between testing and verification. Testing refers to evaluating the system in concrete conditions and making sure that it behaves as expected. Testing has evolved to be a major part of software development ever since its early developments in the 1950s \cite{juran1999quality,myers2011art}. Verification has been defined by ML researchers as "producing a compelling argument that the system will not misbehave under a broad range of circumstances" \cite{goodfellow2018making}. Goodfellow et al. discuss the limitation of existing testing paradigms, highlighting the need for verification frameworks for ML systems as well as the challenges in purely theoretical verification \cite{goodfellow2018making}. We build on their work and propose that verification frameworks need to be able to accommodate the concerns of a broader kinds of actors while also being dynamic. Instead of static code which is evaluated before the system is deployed they need to be able to change and accommodate a dynamic sociotechnical context. The verification scripts which are part of the DAS agreement may be executed at a user endpoint at the same time as they are receiving the algorithmic outcome from a ML system, ensuring that the latest verification script is used.

Cryptography methods such as the Zero-Knowledge algorithm, non-interactive zero-knowledge proofs \cite{blum1988non}, and other privacy-preserving methods, could be utilized in the verification procedures \cite{narula2018zkledger,katz1996handbook}.  

\section{A DASA for a concrete AI model}
Here we develop a DAS agreement for an AI model which is employed in the creation or curation of real and synthetic text, image, video or audio data for the purposes of entertainment, advertising, propaganda or education and the algorithmic targeting of people with this content. For the purposes of the DAS, we assume a scenario where a single user is interacting with an AI model. The model is a Recommender System collaborative filtering model. At each time step, the model aims to show the user a list of recommended items which she might want to engage with (for example reading a news article, watching a movie, acting on a post in social media, etc). 

\begin{itemize}
\item \textbf{Accommodating and enabling change}
Ultimately, the target AI system is trying to model human preferences. A DASA script could employ evaluation metrics which operate on the level of human-algorithmic interactions and measure quantities such as the speed of degeneration of feedback loops \cite{Jiang:2019:DFL:3306618.3314288} or the \textit{barrier-to-exit} metric \cite{rakova2019human}. Further development and adoption of such metrics could provide practitioners with "a proxy to the amount of effort a user needs to expend in order for the system to recognize their change in preference" \cite{rakova2019human}. For example, if the \textit{barrier-to-exit} score goes above a user-defined or predefined threshold the DAS agreement may alarm the user about the observed behavior. This could be an indication of the early formation of effects such as filter bubbles or echo-chambers. 

\item \textbf{Co-constitution}
We argue that the DASA could be a way for different stakeholders to cooperate on identifying and addressing the ethical challenges of RSs \cite{milano2019recommender}. For example, the DASA could provide a means to practically operationalize the well-being impact assessment process proposed by Musikanski et al., broadening the scope of the RS evaluation framework. Musikanski et al. have defined well-being broadly, encompassing the domains of: "(1) Affect, (2) Community, (3) Culture, (4) Education, (5) Economy, (6) Environment, (7) Human Settlements, (8) Health, (9) Government, (10) Psychological Well-Being/Mental well-being, (11) Satisfaction with life, and (12) Work" \cite{p7010}. In the context of RSs, it may be beyond the capabilities of a single organization to adequately address all of these aspects in an iterative manner throughout the whole pipeline of design, development, deployment, etc. However, different stakeholders could contribute to a DASA in a distributed way, while also protecting their intellectual property. For example, in a setting where a RS model is employed by a media platform, a civil rights organization may develop DASA scripts which implement indicators included in the Human Settlements and Government domains. Similarly, social scientists and journalists may contribute other DASA scripts which evaluate indicators within the domain of Satisfaction with life.


Furthermore, a DASA is not limited to a single AI model but also recognizes the second order effects of our interactions with AI models. There are ripple effects of our interactions with RSs - our data goes on to be used in other settings as raw data for other algorithms. A DASA script might try to tackle the quantification of such second order ripple effects.

\item \textbf{Reflexive directionality}
A DASA script could use the list of recommended items as well as other datasets available on the Internet and try to infer different properties of the input data. For example, a verification script might check if the original input data which the user provided had been properly de-identified when it was used within the AI model. To do that, the verification script could use the algorithmic output and try to restore the identity of the user who is receiving the recommendation. The script is being executed at the user endpoint and therefore has the ground truth data of the actual user identity. Recent research and law suits in the healthcare space have shown that it is not always straightforward what de-identification means in certain contexts \cite{articleRajkomar,ggLAwsuit}. Therefore we see an urgent need for broader multi-stakeholder verification frameworks such as the DASA proposal.

\item \textbf{Allow for friction}
A DASA script could provide an explanation of the algorithmic outcome. The design interface could allow the user to explore why did the system generate a specific recommendation vs another. Given that the DASA script is being developed by the RS creator, it could include a feedback loop which allows users to ask and act on questions such as: \textit{Could I correct the course of the algorithm by nudging it into the right direction, most aligned with my true human preferences?} 

\item \textbf{Generativity}
For example, a DASA script written by a school board could have additional verification checks, ensuring that a RS content used in educational contexts is aligned with the age group of their students. The DAS agreement is a set of verification scripts which are modular in nature. Therefore, they could be easily combined and recombined in multiple ways accommodating the specific needs of individuals as well as different communities such as cities, neighbourhoods, communities of common interests, and others. Which scripts are being included in the DASA could be part of the user interface between people and AI systems and in this way be fully transparent. The goal is to provide means for different levels of verification and negotiation between people and AI systems. Users may choose to explore and interact with the DASA scripts at their own will. For example, a user of the RS model may engage and customize the DASA when they are not satisfied with the recommendations being made to them and they are curious to figure out why. On the other hand, users may choose to proactively make sure the RS is aligned with their human preferences before deciding to interact with it.
\end{itemize}

\section{Conclusion and Future Work}
We hope that our proposal on the importance of introducing a Dynamic Algorithmic Service Agreement framework in the context of AI Systems, will inspire tangible discussions and further exploration of how such frameworks could improve the human-algorithmic interactions while upholding societal values for the individual. Future work will develop prototypes in simulated and real-world settings as well as conduct research exploring how could DASAs be practically implemented at scale given the variety of AI models and organizations building and using third party algorithmic models. 
Future research directions will also include discussions on oversight and how such agreements might fit into existing and future legal frameworks.

\bibliography{library}
\bibliographystyle{aaai}

\end{document}